\documentclass[useAMS,usenatbib]{mn2e} 
\usepackage{aas_macros}
\usepackage{graphics}
\usepackage[pdftex]{graphicx}
\usepackage{epstopdf}
\usepackage{epsfig}  
\usepackage{natbib} 
\usepackage{float}
\usepackage{amsmath}
\usepackage{times}
\usepackage[varg]{txfonts}
\usepackage{verbatim} 
\usepackage{multirow,bigdelim} 
\usepackage{color}
\usepackage{vmargin}
\setmarginsrb{1.2cm}{1cm}{1.2cm}{1cm}{1cm}{1cm}{1cm}{1cm}

\newcommand{\hmsun}{{\,\rm h^{-1}M}_\odot}

  \makeatletter
    \renewcommand{\paragraph}{\@startsection{paragraph}{4}{\z@}%
      {-3.25ex\@plus -1ex \@minus -.2ex}%
      {1.5ex \@plus .2ex}%
      {\normalfont\small\centering}}
     
    \renewcommand{\subparagraph}{\@startsection{subparagraph}{5}{\z@}%
      {-3.25ex\@plus -1ex \@minus -.2ex}%
      {1.5ex \@plus .2ex}%
      {\normalfont\small\centering}}
    \makeatother

\newcommand{\ginnungagap}{{\sc Ginnungagap}}

\newcommand{\kms}{{ km~s$^{-1}$}}
\newcommand{\hMpc}{{ h$^{-1}$~Mpc}}

\title[Virgo's formation]{How did the Virgo cluster form?}
\author[Sorce et al.]
{
Jenny G. Sorce$^{1}$\thanks{E-mail: \text{jsorce@aip.de}}, 
Stefan Gottl\"{o}ber$^1$,
Yehuda Hoffman$^{2}$,
Gustavo Yepes$^{3,4}$\\
$^1$Leibniz-Institut f\"{u}r Astrophysik, 14482 Potsdam, Germany\\
$^2$Racah Institute of Physics, Hebrew University, 91904 Jerusalem, Israel\\
$^3$ Departamento de F\'{\i}sica Te\'orica, Universidad Aut\'onoma de Madrid, Cantoblanco, 28049 Madrid, Spain \\
$^4$ ASTRO-UAM, UAM, Unidad Asociada CSIC \\
}

\begin{document}

\date{}


\maketitle

\label{firstpage}

\begin{abstract}

\indent While the Virgo cluster is the nearest galaxy cluster and therefore the best observed one, little is known about its formation history. In this paper, a set of cosmological simulations that resemble the Local Universe is used to shed the first light on this mystery. The initial conditions for these simulations are constrained with galaxy peculiar velocities of the second catalog of the Cosmicflows project using algorithms developed within the Constrained Local UniversE Simulation project. Boxes of 500 \hMpc\ on a side are set to run a series of dark matter only constrained simulations. In each simulation, a unique dark matter halo can be reliably identified as Virgo's counterpart. The properties of these Virgo halos are in agreement at a 10-20\% level with the global properties of the observed Virgo cluster. Their zero-velocity masses agree at one-sigma with the observational mass estimate. In all the simulations, the matter falls onto the Virgo objects along a preferential direction that corresponds to the observational filament and the slowest direction of collapse. A study of the mass accretion history of the Virgo candidates reveals the most likely formation history of the Virgo cluster,  namely a quiet accretion over the last 7 Gigayears.
\end{abstract}


\section{Introduction}

Galaxy clusters are the largest gravitationally bound objects that can be found in the Universe today. They have been largely studied from both observational and theoretical points of view \citep[for an overview, see e.g.][]{1995lssu.conf..209B,2011ASL.....4..204B,2012ARA&A..50..353K}. Even though clusters contain only a small percentage of the galaxies in the Universe, they are powerful probes to study Large Scale Structure, dark matter, galaxy formation and cosmology as a whole when associated with other probes. Among these objects, the Virgo cluster is the closest to us and as such has been studied and observed  in great detail. Thus, it is probably the best known cluster \citep[e.g.][for a non-extensive list]{2000eaa..bookE1822B,2009eimw.confE..66W,2011MNRAS.416.1996R,2012A&A...543A..33V,2011MNRAS.416.1983R,2011ASPC..446...77F,2012ApJS..200....4F,2012MNRAS.423..787T,2014MNRAS.442.2826C,2014ApJ...782....4K,2015A&A...573A.129P,2015ApJ...807...88G}. Its formation history however, and more particularly its assembly history, is  not directly accessible via observations. It can only be inferred indirectly. Therefore cosmological simulations constitute an ideal tool to provide some insights into the cluster evolution. \\

Large volume cosmological simulations  can provide a substantial sample of Virgo candidates selected  in a mass range compatible with observational mass estimates of Virgo. However, because the environment plays an important role in the formation of cosmic objects \citep{2005ApJ...634...51A,2007ApJ...654...53M}, the formation  histories of these Virgo candidates would span over a wide range of possibilities, known as cosmic variance. To reduce this variance, two approaches are available: 1) studies of a very large sample of candidates, maximizing the number of constraint-parameters mimicking the observed Virgo cluster and its environment, to disentangle the distinctive features of clusters like Virgo from the other clusters. It requires lots of knowledge regarding selecting Virgo-like objects that is not necessarily available ; 2) studies of candidates in the proper local large scale environment, using constrained simulations that resemble the Local Universe down to a few megaparsecs, to identify the characteristics of typical counterparts.

This paper is based on the latter path, i.e. on dark matter simulations obtained with constrained initial conditions built within the context of the Constrained Local UniversE Simulations (CLUES) project\footnote{\tt http://clues-project.org/} \citep{2010arXiv1005.2687G}. Unlike common cosmological simulations that starts from random gaussian density fluctuations, the CLUES initial conditions follow a set of observational constraints, here,  from the Cosmicflows project\footnote{http://www.ipnl.in2p3.fr/projet/cosmicflows/} \citep[e.g.][]{2012ApJ...744...43C,2012ApJ...749..174C,2012ApJ...749...78T,2012ApJ...758L..12S,2012AJ....144..133S,2013ApJ...765...94S,2013AJ....146...86T,2014Natur.513...71T,2014MNRAS.444..527S}.  50 \% of these constraints are within 61 \hMpc\ therefore Virgo's region and its neighborhood are very well reproduced.

Actually, these constrained simulations resemble the Local Universe as observed today in terms of the Large Scale Structure (superclusters, voids, etc) down to a few megaparsecs  \citep{2016MNRAS.455.2078S}. Each one of these simulations contains a dark matter halo at a similar position and with a similar mass as the observed Virgo cluster. These halos constitute our Virgo candidates whose properties and evolution are studied in details.\\

More precisely, the paper is structured as follows. Section 2 gives a general overview on the constrained simulations. In Section 3, in each simulation the unique Virgo halo at redshift zero is identified and compared with the observed Virgo cluster. Further, the environment of these halos up to a few virial radii is studied, providing a total mass that includes the outskirts of the halos and highlighting a preferred direction of infall.  Finally the formation histories of the Virgo halos compared to merging histories of random halos sharing the same mass are presented.


\section{Constrained Simulations of the Local Universe}

The second generation observational catalog of radial peculiar velocities, built by the Cosmicflows collaboration and used as constraints within the CLUES project, has already been widely described in \citet{2013AJ....146...86T}. Briefly, it contains more than 8,000 accurate galaxy peculiar velocities derived with distance measurements obtained mostly from the Tully-Fisher relation \citep{1977A&A....54..661T} and the Fundamental Plane methods \citep{2001MNRAS.321..277C}. The rest comes from Cepheids \citep{2001ApJ...553...47F}, Tip of the Red Giant Branch \citep{1993ApJ...417..553L}, Surface Brightness Fluctuation \citep{2001ApJ...546..681T}, supernovae of type Ia \citep{2007ApJ...659..122J} and other miscellaneous methods. 

Since constrained initial conditions can be constructed only above the scale of non-linear motions, we use the grouped version (552 groups and 4303 single galaxies) of the catalog in which all local motions within clusters or groups are removed  \citep[e.g.][]{2015AJ....149...54T,2015AJ....149..171T}. This grouping also reduces the uncertainties on measurements, nonetheless, \citet{2013AJ....146...86T} warns us that the catalog must be handle with care because of residual biases. An iterative method, explained in \citet{2015MNRAS.450.2644S}, to minimize the spurious infall onto the Local Volume and to reduce spurious non-gaussianities in the radial peculiar velocity distribution has also been applied to obtain a new distribution of radial peculiar velocities and corresponding distances.

The constrained initial conditions are constructed from this grouped and corrected dataset of radial peculiar velocities as described  in \cite{2014MNRAS.437.3586S,2016MNRAS.455.2078S}. Here, the four steps are summarized and briefly described to highlight their purposes:
\begin{itemize}
\item The Wiener-Filter method \citep[WF,][]{1995ApJ...449..446Z,1999ApJ...520..413Z} reconstructs the cosmic displacement field required to account for the displacement of constraints from their precursors' locations, 
\item the Reverse Zel'dovich Approximation (RZA) relocates constraints at the positions of their progenitors \citep{2013MNRAS.430..888D,2013MNRAS.430..902D,2013MNRAS.430..912D} and replaces noisy radial peculiar velocities by their 3D reconstructions \citep{2014MNRAS.437.3586S},
\item the Constrained Realization \citep[CR,][]{1991ApJ...380L...5H,1992ApJ...384..448H} of Gaussian field technique produces density fields constrained by observational data, adding a random realization (RR) to have power in agreement with the power spectrum of the underlying model. These latter are then converted into a white noise field.
\item In order to increase the resolution random small scale features are added\footnote{Here the \ginnungagap\ code, https://github.com/ginnungagapgroup/ginnungagap, is used.}, up to the required level, to the white noise field. Finally the resulting white noise is Fourier transformed and convolved with the power spectrum to build the initial conditions for cosmological simulations.
\end{itemize}

Fig. 1 shows the power spectra of 15 realizations of initial conditions.  There is a small change compared to the process described  in  \citet{2016MNRAS.455.2078S}.  Guided by tests with mock catalogues obtained from a large unconstrained simulation, the treatment of errors has been improved. Neither the conclusions of \citet{2016MNRAS.455.2078S} nor those of this paper are altered, in particular it does not change the properties of Virgo halos.

 \begin{figure}
\hspace{-0.5cm}\includegraphics[width=0.5 \textwidth]{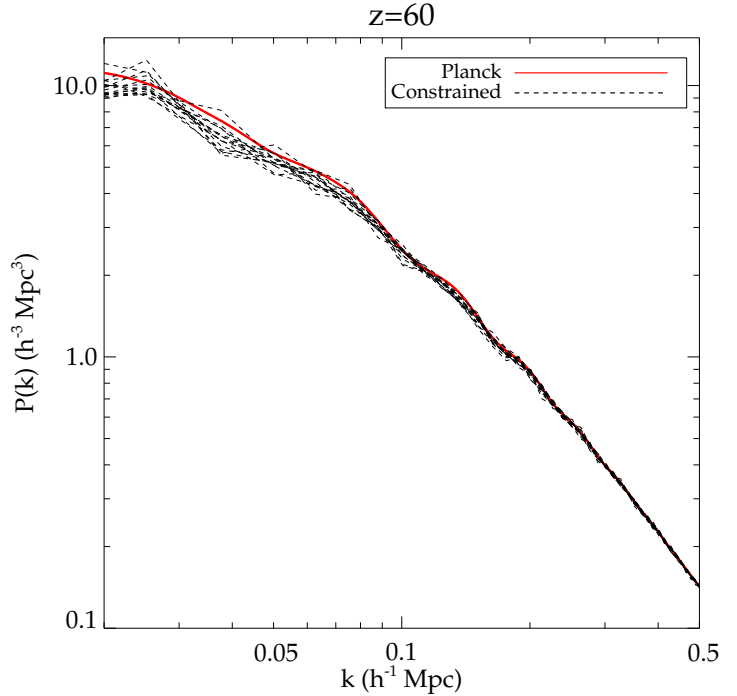} 
\caption{Power spectra of 15 constrained initial conditions at z=60 (dashed black lines) and the Planck power spectrum (solid red line).}
\label{fig:powspec}
\end{figure}
 
Using the technique previously described,  a set of 25 constrained simulations with 512$^3$ particles and a box size of 500 \hMpc\ have been performed.  Fifteen of these simulations are done based on different random realizations of gaussian fields. They are called hereafter different-RR simulations. The remaining ten simulations share the same random large scale field but different small scale features have been added to increase the resolution \citep[namely, we repeated the procedure described in detail in][]{2016MNRAS.455.2078S}. From now on, they are referred to the same-RR simulations. The two subsets allow us to evaluate to which extent the large and the small (non-linear and thus unconstrained) scales influence the evolution and formation history. Two simulations with 1024$^3$ particles with the same box size have been run to check that results shown in this paper are not  affected by mass resolution\footnote{A halo of  2$\times$10$^{14}$ $\hmsun$  (the lowest mass estimate of the Virgo halos is 2.7$\times$10$^{14}$ $\hmsun$)  consists of  2,500 particles, a typical progenitor at z=2 contains still more than 200 particles having  512$^3$ particles within the box.}. Simulations are run within the framework of Planck cosmology \citep[$\Omega_m$=0.307, $\Omega_\Lambda$=0.693, H$_0$=67.77, $\sigma_8=0.829$,][]{2014A&A...571A..16P}. The starting redshift is z=60 and the force resolution is set to 25 $h^{-1}$ kpc.


\section{Virgo Halos}

\begin{table*}
\begin{center}
\begin{tabular}{ccccccccccc}
\hline
\hline
Cluster & Dist & sgl & sgb & sgx & sgy & sgz& v & R$_{\rm ZV}$ &  M$_{\rm ZV}$ & $\sigma_v$ \\
Virgo & 11.1 & 103.2 & -2.612 & -2.56 &10.9 & -0.512 & 645 & 4.2 $\pm$ 0.8$^a$ - 4.9 $\pm$ 0.5$^b$ & 3.9 $\pm$ 2.1 $^a$ - 5.4 $\pm$ 1.6 $^b$& 665  \\
\hline
\hline
\end{tabular}
\end{center}
\caption{Observational Parameters of the Virgo Cluster: (1) Cluster name, (2) Distance to the Milky-Way in \hMpc\ from \citet{2013AJ....146...86T},  (3) and (4) supergalactic longitude and latitude coordinates, (5) to (7) x, y and z supergalactic coordinates in \hMpc, (8) reconstructed 3D velocity with respect to the Cosmic Microwave Background in \kms, (9) Zero-velocity radius  and the one-sigma uncertainty interval in \hMpc, (10) Mass derived from the zero-velocity radius and the one-sigma uncertainty (10$^{14}$ h$^{-1}$ M$_\odot$) from \citet{2011A&A...532A.104N,2010MNRAS.405.1075K}$^a$ and \citet{2014ApJ...782....4K}$^b$, (11) velocity dispersion from \citet{2013AJ....146...86T} and the Extragalactic Distance Database \citep[http://edd.ifa.hawaii.edu,][]{2009AJ....138..323T} in \kms.}
\label{Tbl:2}
\end{table*}

The Virgo cluster has been widely studied (see for example the non-extensive list given in the introduction section) and some of its characteristics are now well determined and given in Table \ref{Tbl:2}. In order to compare these observational properties with the simulated ones, we establish for each simulation a list of halos and their properties using Amiga Halo Finder \citep[AHF,][]{2009ApJS..182..608K} and the definition based on M$_{\rm 200}$ ( i.e the mass enclosed in a sphere with a mean density of 200 times the critical density of the Universe). 

The Virgo counterparts in each simulation are then identified, from the AHF catalogs, as follows. By construction of the constrained simulations,  the observer is always located  in the center of the box. The coordinate system in the simulations is defined to match the observational supergalactic coordinates. Then, spherical regions of 5 \hMpc\ radius centered on the observed position of Virgo are analyzed.  In each simulation, in this spherical region, a dark matter halo of reasonable mass (same order of magnitude), namely a unique Virgo candidate at the right place, is found. The characteristics of these halos are compared with those of the observed Virgo in the next subsection. 

\subsection{General properties at redshift zero}

\begin{figure}
\includegraphics[width=0.48 \textwidth]{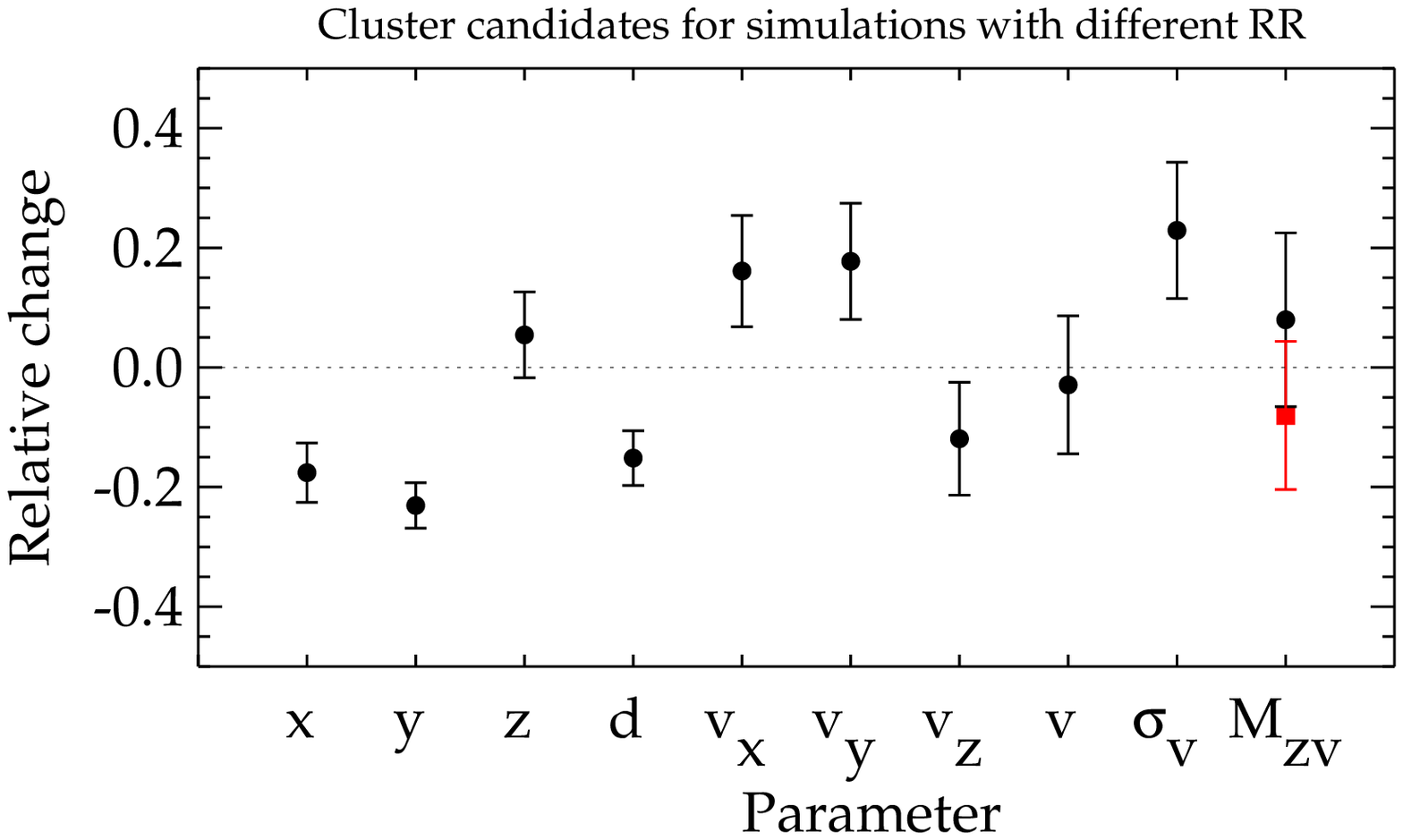} \\

\vspace{0.2cm}
\includegraphics[width=0.48 \textwidth]{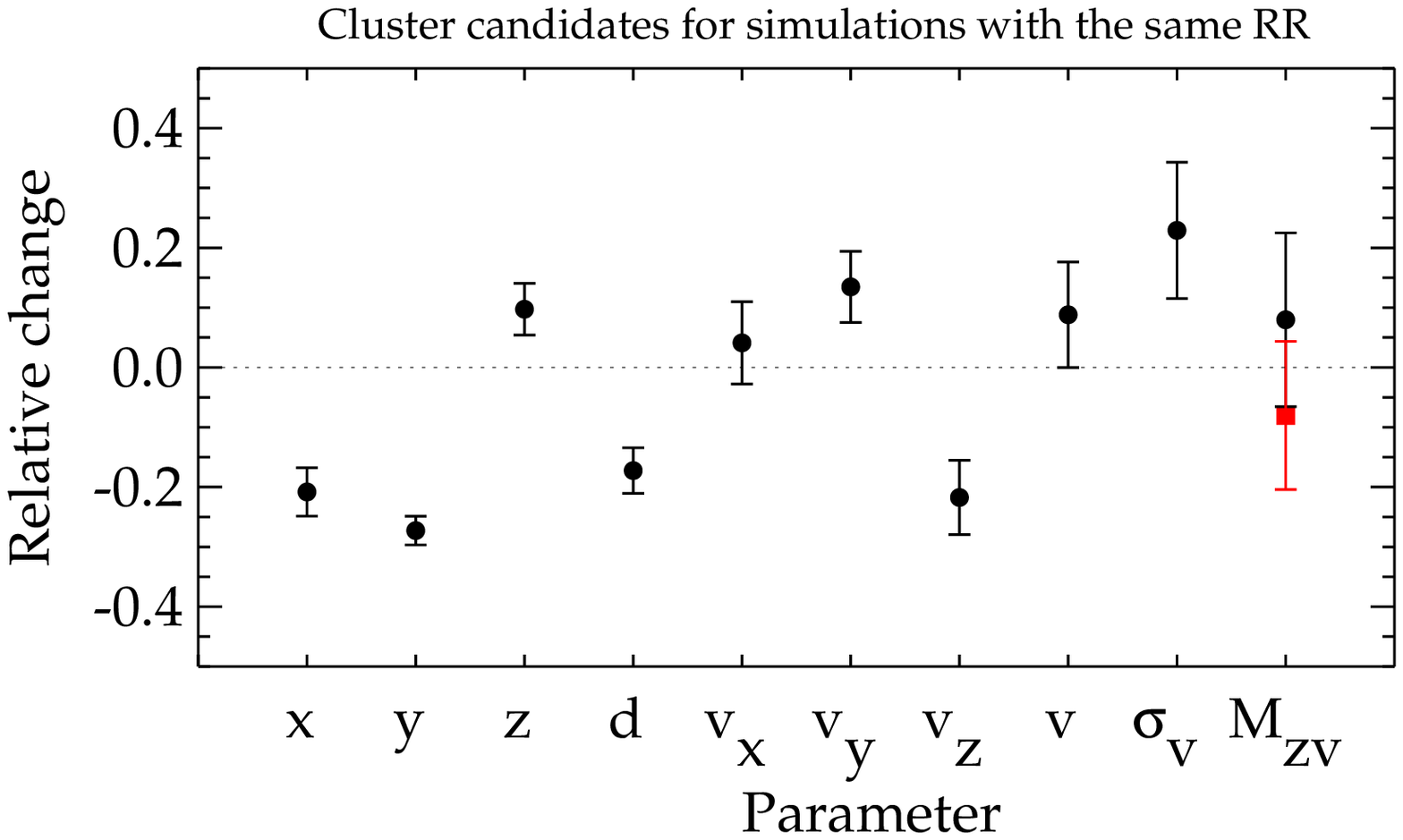} 
\caption{Relative change of the parameters of simulated Virgo halos with respect to that of the observed Virgo cluster. The relative change  is defined as the difference   between the simulated and the observed,  or reconstructed,  component divided  by the corresponding observational quantity. Namely, the distance for the supergalactic coordinates and the distance itself, the reconstructed 3D velocity for the velocity components and the velocity vector, the measured velocity dispersion for the velocity dispersions and the estimated mass (black, mean zero-velocity mass estimate - red, the most recent zero-velocity mass estimate) of the observed cluster for the masses. Top: Virgo halos in the fifteen constrained simulations based on different large scale random realization fields. Bottom: Virgo halos in the ten constrained simulations sharing the same large scale random realization field but different small scale features.}
\label{fig:Virgorelatchange}
\end{figure}

In order to estimate the agreement between simulated Virgo halos and the Virgo cluster, the relative change of  coordinates, velocities, velocity dispersions and masses are considered.  We define this quantity as the difference between the simulated and the observed or reconstructed component divided by the observational: 1) distance for the coordinates and distances, 2) 3D reconstructed velocity for the velocities and their components, 3) velocity dispersion itself for velocity dispersions and 4) (mean) mass estimate for the masses.ÕAs a rough approximation, dark matter particles and their velocities in the simulation are directly compared to galaxies and their velocities. Figure \ref{fig:Virgorelatchange} shows the mean and  the one-sigma scatter of the relative changes of the parameters, defined above, for the Virgo halos found in the different constrained simulations.  The top panel shows the mean relative changes (filled black circles) and their standard deviations (black error bars) for the Virgo halos in the 15 different-RR simulations while the bottom panel gives those of Virgo halos found in the 10 same-RR simulations.

Relative changes in the different quantities are low (around 10-20\% on average) indicating that observed-reconstructed and simulated properties of Virgos are similar. Note that standard deviations between the different simulations are also quite small (again about 10-20\%), indicating an effective  reduction of the cosmic variance. As expected, the 10 same-RR simulations host Virgo halos in slightly better agreement with each other (scatter less than 5-10\% of the considered parameter). 

Masses to be compared in detail (i.e. not just by order of magnitude) with the observational mass estimates are not direct outputs of the Halo Finder. Indeed, e.g. M$_{\rm 200}$ masses are known to miss more than 25\% of the mass when compared to friends-of-friends algorithms \citep[e.g.][]{2011MNRAS.414.3166A}. As for virial masses M$_{\rm vir}$, their observational counterparts can suffer from great uncertainties due to outliers included in the galaxy system by projection effect \citep{1990Ap&SS.170..347P,1990A&A...237..319P} and/or, an absence of virial equilibrium because of merging or accretion episodes \citep{2002ASSL..272...39G}.

An observer's scheme consists in estimating the zero-velocity radius - distance where radial velocities relative to the center of mass are zero - to derive a mass estimate for the clusters. From an observational point of view, zero-velocity radii provide mass estimates totally independent from the virial mass estimates. Applying the same scheme to dark matter halos allows comparisons between masses obtained with the same method in both observations and simulations.

\subsection{Mass estimate}

\begin{figure}
\hspace{-0.5cm}\includegraphics[width=0.51 \textwidth]{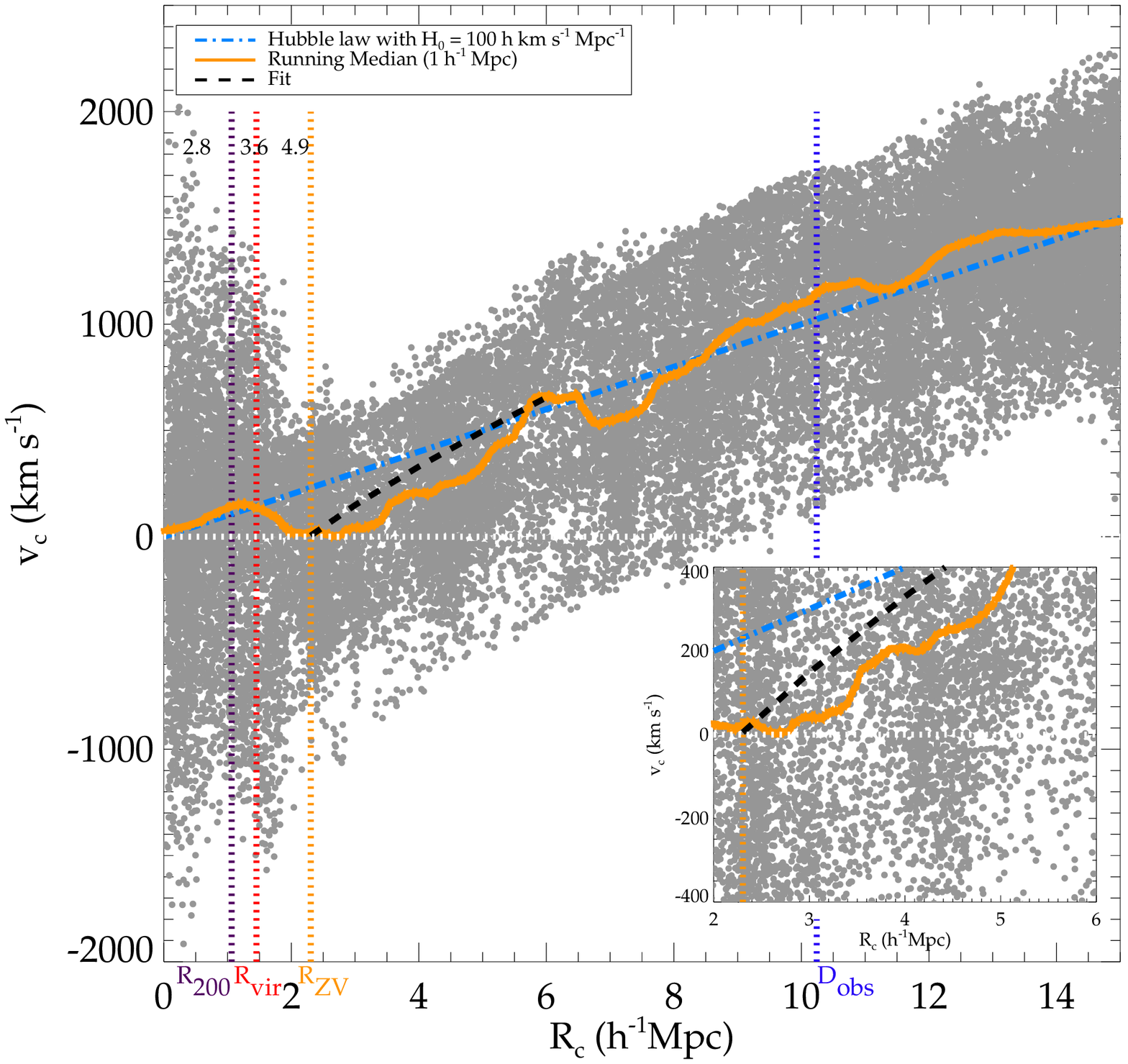}\\

\hspace{-0.6cm}\includegraphics[width=0.515 \textwidth]{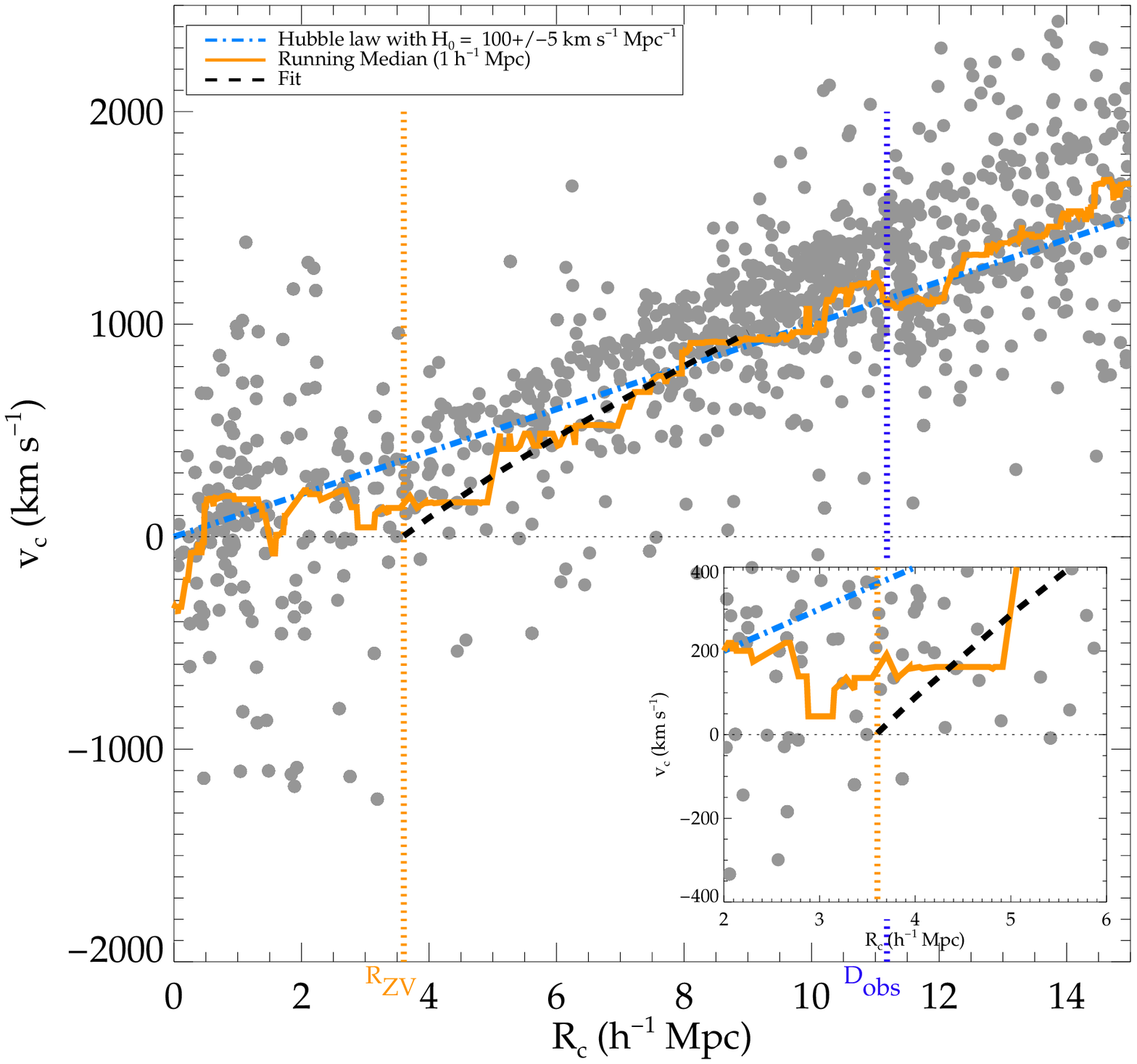}
\caption{Hubble diagram for one Virgo dark matter halo (top) and for the observed Virgo cluster with galaxies from the second catalog of the Cosmicflows project. Grey dots stand for the particles respectively for the galaxies. The running median with a 1 \hMpc\ window is plotted as a solid yellow line using every particles/galaxies. The straight dotted-dashed light blue line stands for the Hubble Law with H$_0$ = 100 h \kms Mpc$^{-1}$. The dashed black line stands for the spherical collapse fit. Lines have values equal to R$_{\rm 200}$ (violet dotted line), virial (red dotted line only in the top panel) and zero-velocity (yellow dotted line) radii. The thick dotted blue line stands for the distance to the center of the box/us respectively. In the top panel, M$_{\rm 200}$ and M$_{\rm vir}$, in 10$^{14}$ $\hmsun$ unit, are written on the left side below the legend. Next to them, the mass obtained using all the particles within the zero-velocity radius is given using the same system unit. The small bottom right panels are zoom onto the zero-velocity regions.}
\label{fig:outflow}
\end{figure}

\begin{figure}
\hspace{-0.5cm}\includegraphics[width=0.5 \textwidth]{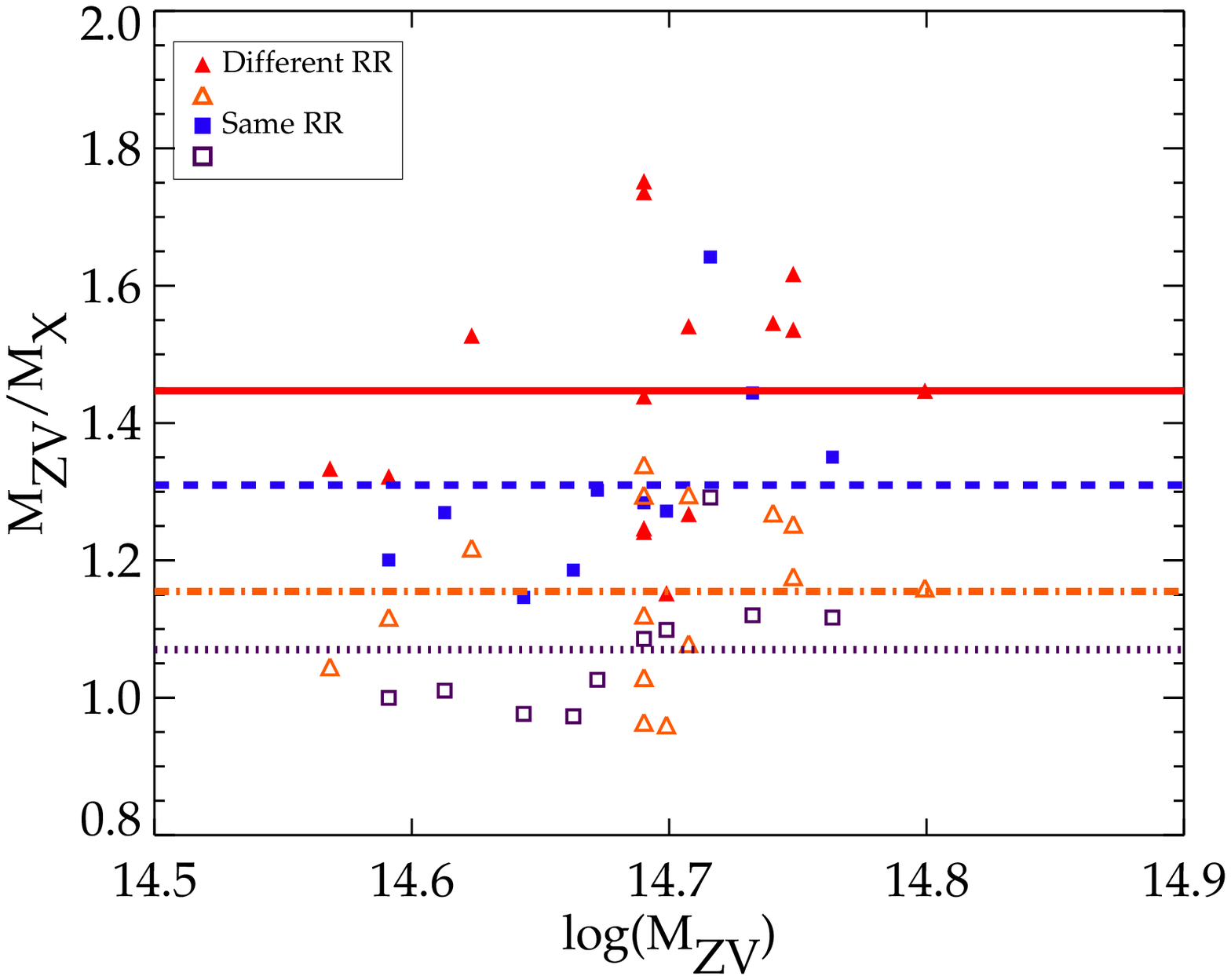}
\caption{Ratios of zero velocity mass estimates to M$_{\rm 200}$ or M$_{\rm vir}$ for the 15 Virgo halos obtained in the 15 constrained simulations build with different large scale random field simulations (red filled triangles or orange triangles) and for the 10 Virgo halos in the 10 constrained simulations sharing the same large scale random realization (blue filled squares or violet squares). The solid red line (resp. orange dot-dashed line) shows the mean of the ratios M$_{\rm ZV}$/M$_{\rm 200}$ (resp. M$_{\rm ZV}$/M$_{\rm vir}$) for the 15 Virgo halos while the dashed blue line (resp. violet dotted line) shows the mean  M$_{\rm ZV}$/M$_{\rm 200}$ ratio (resp. M$_{\rm ZV}$/M$_{\rm vir}$ ratio) for the 10 Virgo halos.}
\label{fig:massratio}
\end{figure}

In this subsection only, we consider the full velocity of the particles, i.e. including the Hubble flow. 
Then,velocities v$_c$ and distances R$_c$ of each particle in each Virgo halo-centric system are derived with the formula and geometric considerations given by \citet{2006Ap.....49....3K} and described in the Appendix. Figure \ref{fig:outflow} presents one of the resulting Hubble diagrams (top) and that of the observed Virgo cluster (bottom). Note that in the case of the observed Virgo cluster, galaxy distance measurements, and by extension velocities, are subject to uncertainties that are not considered here, the diagram is shown only for overall comparison purposes. For an extensive study of the Hubble diagram of the observed Virgo cluster with error bars, the reader should refer to \citet{2014ApJ...782....4K}. In addition, R$_{\rm 200}$ and the virial radius R$_{\rm vir}$, as given by the Halo Finder, are plotted and the total masses (in 10$^{14}$ $\hmsun$ unit) of the particles encompassed in these regions are given on the left side of the top diagram below the legend. Next running medians with a window of 1 \hMpc\ are calculated and plotted on top of the two diagrams.

The zero-velocity radius R$_{\rm ZV}$ is the distance where radial velocities relative to the center of mass have null values. By definition, it is a priori the intersection of the running median with the x-axis. Because of a high velocity dispersion in clusters' centers, running medians are, however, noisier close to the centers than in the far outskirts where the Hubble flow is reached. This complicates the estimation of the zero-velocity radii, especially for the observations where uncertainties and low statistics render the signal very noisy. Consequently, to find zero-velocity radii, a theoretical velocity profile derived from the spherical model by \citet{2008A&A...488..845P} is generally fit. This profile assumes a spherical collapse model, $\Lambda$ included, in the outskirts of the clusters. Inside the viral radius, where most of the mass is contained and shell crossing  has already happened, the orbits are mainly radial. Then:
\begin{equation}
V(R_c)=1.377 H_0 \times R_c - 0.976 \times \frac{H_0}{R_c^n} \times
\left(\frac{GM}{H_0^2}\right)^{(n+1)/3}
\end{equation} 
with M the halo mass enclosed in the zero velocity radius, R$_c$ the distance of the particle to the center of mass, V(R$_c$)=v$_c$ the radial velocity of the particle with respect to the halo center and n=0.627 at z=0. Because the model is valid only up to the point where the Hubble Flow is reached, fits are based on particles located in the close by outskirt. The process is iterated until the model converges to the data (dashed black line). By definition, V(R$_{\rm ZV}$)=0 and solving this equation gives R$_{\rm ZV}$ shown in Figure \ref{fig:outflow} by a dotted yellow line. The mass M$_{\rm ZV}$ of the halo is then the number of particles within the zero-velocity radius distance times the particle mass.

These mass estimates are compared to zero-velocity observational mass estimates in Figure \ref{fig:Virgorelatchange} (black filled circle for the mean observational zero-velocity estimate and filled red square for the most recent zero-velocity estimate). M$_{\rm ZV}$ masses reach values very close (within one-sigma), if not identical, to observational mass estimates. For information, Figure \ref{fig:massratio} draws a comparison between M$_{\rm ZV}$ masses and those given by the Halo Finder. For instance, M$_{\rm 200}$ masses are increased on average by 45$\pm$18 \% (31$\pm$14 \% when considering the same-RR simulations) and M$_{\rm vir}$ masses are slightly increased on average by 12$\pm$11 \% (11$\pm$10 \% for the same-RR simulations). 

\subsection{Infall at redshift zero: a preferred direction}

\begin{figure}
\hspace{-0.cm}\includegraphics[width=0.48 \textwidth]{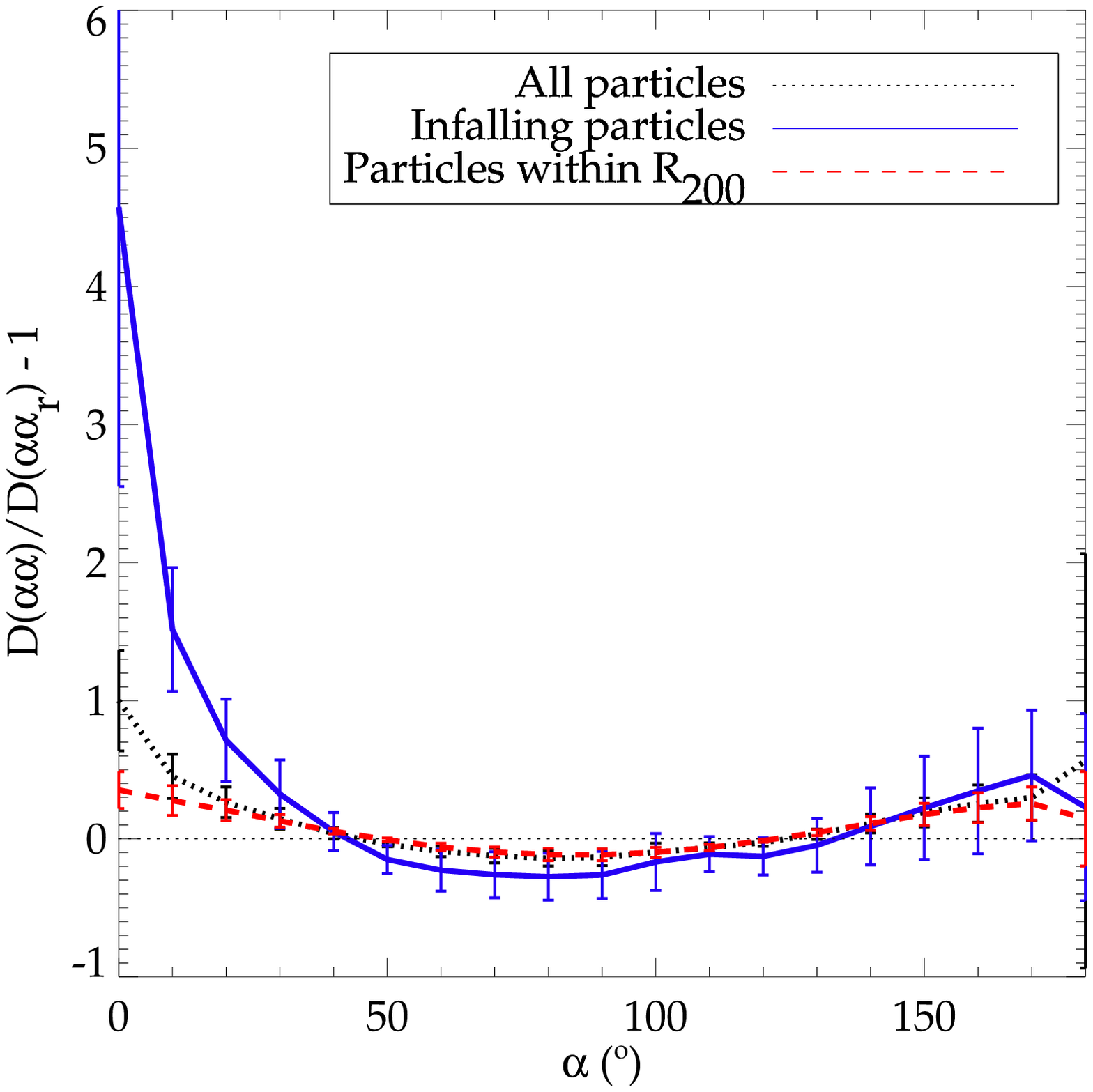}
\vspace{-0.cm}
\caption{Mean angular autocorrelation functions for particles infalling onto the halos (blue solid line), for particles closely bound, i.e. within R$_{\rm 200}$, (red dashed line) and for all the particles within a 6 \hMpc\ sphere centered on the halos (black dotted line). Error bars stand for standard deviations of the mean correlation functions.}
\label{fig:corrfunc}
\end{figure}

\begin{figure*}
\hspace{-0.7cm}\includegraphics[width=1 \textwidth]{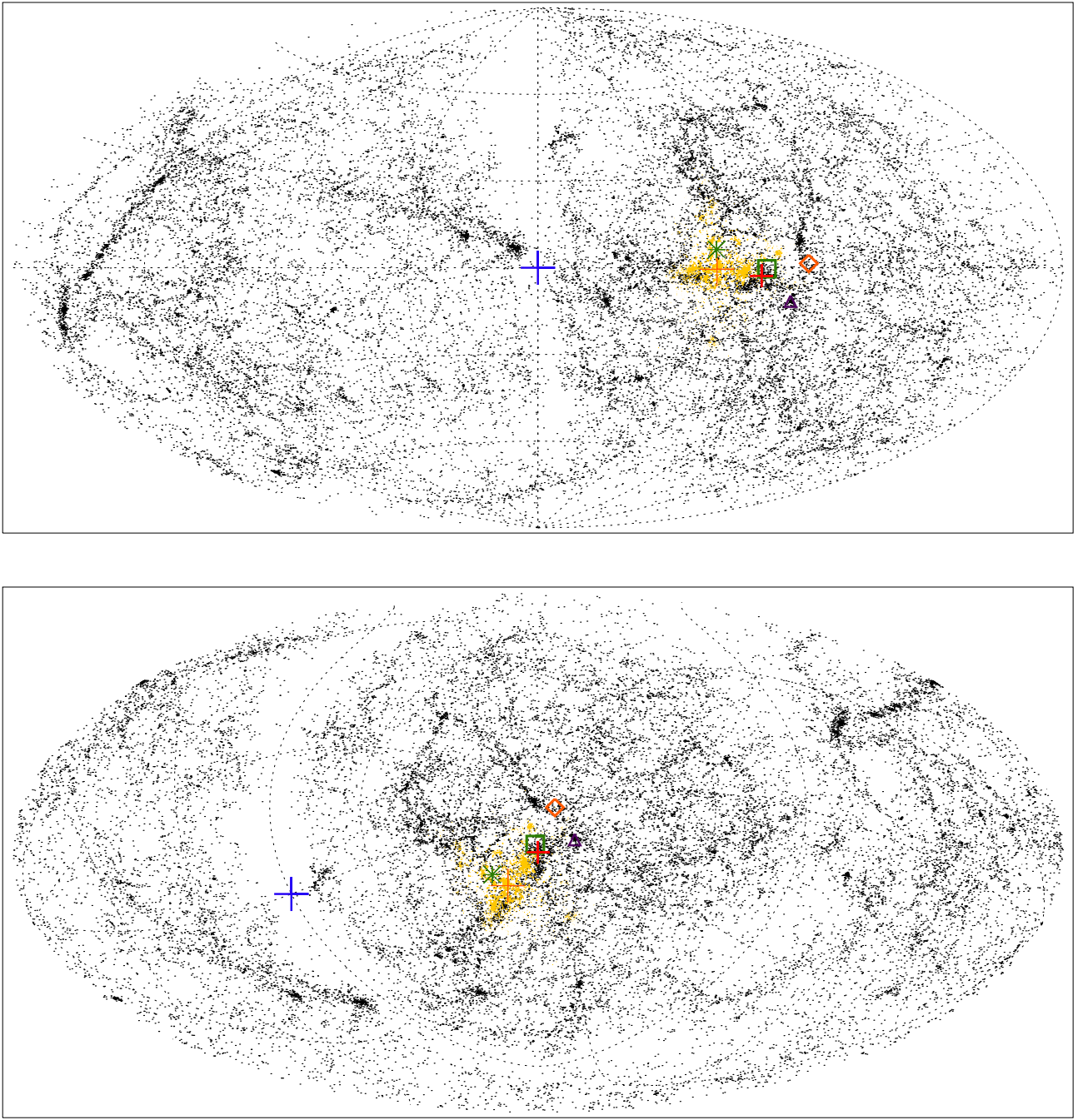}
\vspace{-0.cm}
\caption{Aitoff projection in supergalactic (top) and galactic (bottom) coordinates of galaxies from the V8k redshift catalog (black dots), of the observer (thick blue cross), of the observed Virgo cluster (thick red cross), of one Virgo halo (orange cross) and of the particles infalling onto this halo at redshift zero (yellow dots) The size of the dots in this latter case is proportional to the infall velocity.  The orange diamond together with the orange cross point the principle direction of infall onto the selected for the plot Virgo halo. The green star represents the average position of the Virgo halos and together with the green square they show the average direction of infall. The violet triangle stands for the Abell 1367 cluster. Top: in supergalactic coordinates, centered on the observer. Bottom: in galactic coordinates, centered on the observed Virgo cluster.}
\label{fig:aitoff}
\end{figure*}

\begin{table*}
\begin{center}
\begin{tabular}{cccccccc}
\hline
\hline
Redshift & sgx & sgy & sgz& sgx$_p$ & sgy$_p$ & sgz$_p$ & inclination \\
\hline
2 & 0.93 $\pm$ 0.7 & 5.1 $\pm$ 0.8 & 2.9 $\pm$ 1.0 & 1.5$\pm$ 0.7 & 5.5$\pm$ 0.6 & 3.3 $\pm$ 1.1 & -5$^\circ$\\
0.5 & -2.3 $\pm$ 0.5 & 7.0 $\pm$ 0.3 & 1.2 $\pm$ 0.8 & -2.3 $\pm$ 0.6 & 7.9 $\pm$ 0.5 & 1.6 $\pm$ 1.4 & -3$^\circ$\\
0.25 & -3.3 $\pm$ 0.5 & 7.6 $\pm$ 0.4 & 0.74 $\pm$ 0.8 & -3.2 $\pm$ 0.7 & 7.9 $\pm$ 0.5 & 0.88 $\pm$ 1.0 & -2$^\circ$\\
0 & -4.5 $\pm$ 0.6 & 8.3 $\pm$ 0.4 & 0.99 $\pm$ 0.8 & -4.8 $\pm$ 0.6 & 9.3 $\pm$ 0.5 & 0.26 $\pm$ 0.8 & 5$^\circ$ \\
\hline
\hline
\end{tabular}
\end{center}
\caption{Mean parameters at a given redshift of the particles, selected to be within R$_{\rm 200}$ of Virgo halos at redshift zero, that are infalling onto the progenitors at the same redshift (three first lines) and of the particles within a 6 \hMpc\ radius sphere infalling onto the halo at redshift zero (fourth line): (1) Redshift ; (2) to (4) supergalactic coordinates of the mean center of mass of the progenitors and halos ; (5) to (7) supergalactic coordinates of the mean positions of the infalling particles ; (8) inclination of the preferential direction of infall with respect to the line of sight. The `minus' sign is arbitrary and is here only to show that the infall direction has evolved over time.}
\label{Tbl:3}
\end{table*}

In this subsection, the infall of particles onto the halos at redshift zero is the object of focus and the Hubble expansion is not considered anymore. We restrict ourselves to the 15 different-RR simulations, because the 10 same-RR simulations have, by construction, very similar large scale environment of the Virgo halo. Therefore, using these latter would underestimate the true residual cosmic variance. We aim at studying the (an)isotropy of the infall onto Virgo, and to uncover a preferred direction of infall - if any. Following \citet{2011MNRAS.411.1525L}, an autocorrelation function permits to assess the degree of anisotropy. Defining $\alpha$ as the angle formed by two particles with the center of mass of the halo and $\alpha_r$ as that between a particle and a random point from a random sample with as many points as particles, the autocorrelation function is defined as:
\begin{equation}
\omega(\alpha)=D(\alpha\alpha)/D(\alpha\alpha_r)-1
\end{equation}
where D($\alpha\alpha$) is the distribution of all the angles $\alpha$ and D($\alpha\alpha_r$) that of all the angles $\alpha_r$.

We apply the above scheme and derive the autocorrelation functions for dark matter particles infalling on halos (i.e. at a distance greater than R$_{\rm 200}$ to exclude closely bound particles, with a negative velocity and restricting ourselves to a search radius of  6\hMpc\ from the halo center of mass). The sample of particles within R$_{\rm 200}$ and all the particles within R$_c$=6 \hMpc\ together are also considered. After checking that the result is independent of the considered simulations, results are averaged. The three resulting autocorrelation functions as well as their standard deviations are shown in Figure \ref{fig:corrfunc}.

As expected, dark matter particles within the closely bound regions of the halos do not present any particular arrangement (red dashed line), they are homogeneously distributed around the center of masses. When considering all the particles within a 6 \hMpc\ radius sphere, the layout is not that much different, although the trend of a preferential direction begins to appear more clearly (black dotted line). Finally, selecting only infalling particles outside the closely bound regions but within 6 \hMpc\ the signal (blue solid line) is quite strong: there are 5 $\pm$ 2 times more particles forming a small angle between themselves than when considering a random distribution and the particles, i.e. infalling particles are non uniformly distributed around the halos. Knowing that particles falling onto the inner part of dark matter halos are arranged along a preferred direction, the mean positions of the infalling particles in the different simulations can be derived. We find that the mean coordinates obtained for the different simulations are very similar in all the realizations. The resulting direction of `close' (in the sense that only particles within 6 \hMpc\ are considered) infall is, on average, along the axis formed between the mean position of Virgo halos (-4.5 $\pm$ 0.6 ; 8.3 $\pm$ 0.4 ; 0.99 $\pm$ 0.8) \hMpc\ in supergalactic coordinates and the point of supergalactic cartesian coordinates (-4.8 $\pm$ 0.6 ; 9.3 $\pm$ 0.5 ; 0.26 $\pm$ 0.8) \hMpc. \\

To compare with the observed infall, it seems reasonable to look at farther particles. The resulting `far' infall is illustrated by Aitoff projections in Figure \ref{fig:aitoff} where particles (yellow dots) are shown to fall along the filament linked to the observed Virgo cluster (thick red cross)/ simulated Virgo halo (orange cross). The particles come on average from the supergalactic cartesian coordinates (-1.5,7.5,-0.05) \hMpc, the supergalactic longitude and latitude (101.45,-0.337) in degrees and the galactic longitude and latitude (287.57,76.77) in degrees. The top projection is centered on the observer (blue cross) while the bottom projection is centered on the observed Virgo cluster. This is in agreement with \citet{2000ApJ...543L..27W} who suggest that infall of material along the Virgo-Abell 1367 filament has driven the formation of the Virgo cluster. The observational latitude and longitude coordinates, (235.08,73.02) in degrees, of the Abell 1367 cluster (violet triangle) are indeed very close to the direction of infall in the simulations.\\

Using the concept of the cosmic web, another recent study showed that the infall onto Virgo happens along the direction of slowest collapse \citep{2015MNRAS.452.1052L} corresponding to the filament within which Virgo resides. A number of methods have been developed over the years to derive the cosmic web \citep[e.g.][among others]{2008MNRAS.383.1655S}. Here, the cosmic web is characterized according to the Hessian formalism applied to the velocity field resulting in the shear tensor \citep[][]{2012MNRAS.425.2049H,2012MNRAS.421L.137L}. Note that applied to the gravitational potential, it would give the tidal tensor \citep{2007MNRAS.381...41H}. Detailed equations are given in \citet{2014MNRAS.441.1974L} and only the adopted definition of the shear tensor $\Sigma$ is reminded:
\begin{equation}
\Sigma_{ij}=-\frac{1}{2H_0}(\frac{\partial v_i}{\partial r_j}+\frac{\partial v_j}{\partial r_i})
\end{equation}
where i, j are $x$, $y$ or $z$ and H$_0$ is the Hubble Constant, $v$ is the velocity field and $r$ the location vector. This definition, applied to the linear reconstruction of the Local Universe obtained with the second catalog of the Cosmicflows project, smoothed at 5 \hMpc, gives, at Virgo's location, the three following eigenvectors, ordered by direction of fastest to slowest collapse \textbf{e1}=(0.113,0.956,-0.269) ; \textbf{e2}=(0.167,0.248,0.954) ; \textbf{e3}=(-0.979,0.153,0.131) with the eigenvalues (0.324,0.242,-0.0350). These eigenvectors represent the smooth mean values in the Local Universe at Virgo's position. The third eigenvector, i.e. the direction of slowest collapse, is indeed aligned with the mean direction of `far' infall since the cosine of the angle formed by this third eigenvector and the direction of `far' infall is 0.97.  This corroborates the claims from \citet{2015MNRAS.452.1052L} that the infall onto the Virgo cluster happens along the direction of slowest collapse that also corresponds to the filament to which Virgo belongs.\\
 
 
\subsection{Infall at higher redshifts: Early formation of a filament}

\begin{figure}
\hspace{-0.cm}\includegraphics[width=0.45 \textwidth]{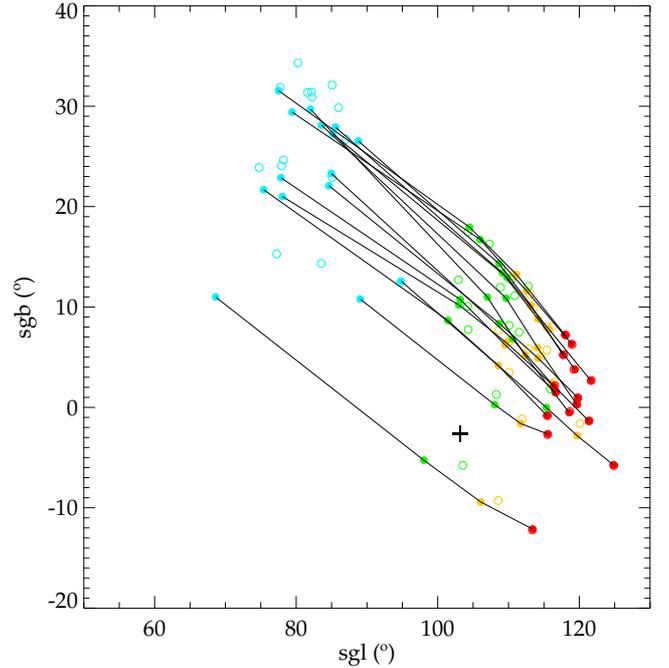}
\vspace{-0.1cm}
\caption{Mean (filled circles) and median (open circles) supergalactic latitudes and longitudes of particles, that are within the R$_{\rm 200}$ radii of the Virgo halos at redshift zero, assuming an observer at the center of the box. The positions are mean positions at different redshifts of the considered particles: 2 (light blue), 0.5 (light green), 0.25 (yellow) and 0 (red) and the solid black lines link the mean positions at different redshifts. The black cross marks the position of the observed Virgo cluster.}
\label{fig:sglsgb}
\end{figure}

We consider the particles closely bound into Virgo halos at redshift zero (i.e. particles within R$_{\rm 200}$) and we trace them back to obtain their positions at earlier redshifts: 2, 0.5, 0.25. Figure \ref{fig:sglsgb} shows the mean positions of the selected particles at these three different redshifts. This figure gives us a visual appraisal of the motion of the center of mass of Virgo dark matter halos from earlier time up to today. An agreement between the simulations, apart for three (reduced to two at lower redshifts) slightly outliers in positions but not in direction of motion, at different redshifts is clearly visible.

\begin{figure*}
\includegraphics[width=0.9 \textwidth]{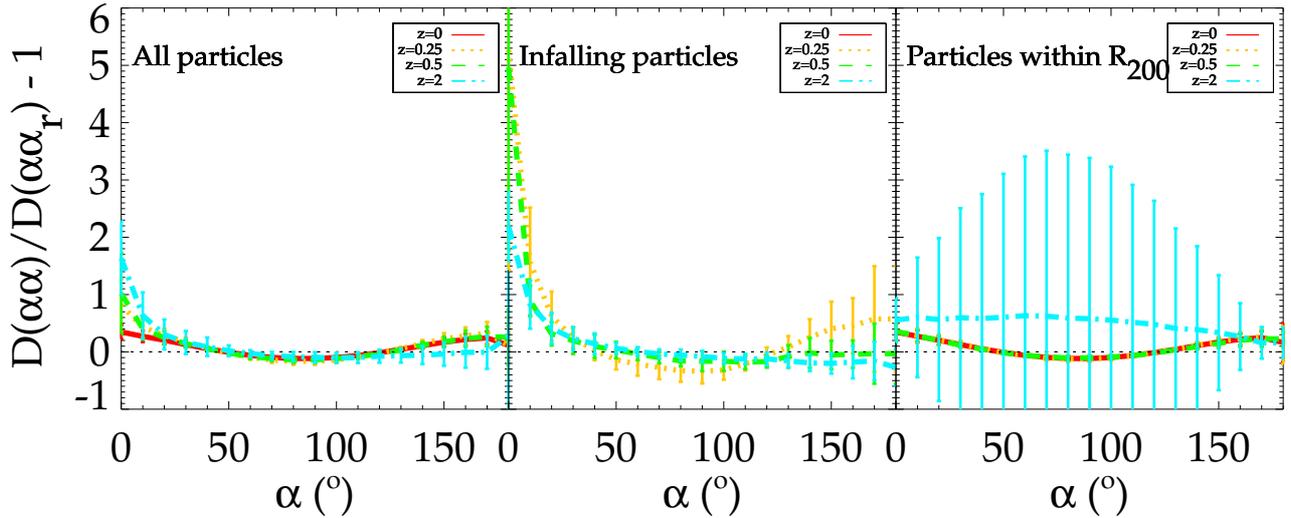}
\caption{Mean angular autocorrelation functions at different redshifts computed for the particles that are closely bound at redshift zero (i.e. within the R$_{\rm 200}$ radius sphere of the Virgo halos at $z=0$) tracing back their positions at earlier redshifts. Left panel: all the selected particles using their positions at a given redshift. Middle panel: particles beyond R$_{\rm 200}$ of the halo progenitors at a given redshift which are infalling onto the progenitors at the same redshift. Right panel: particles within R$_{\rm 200}$ radii of the Virgo progenitors. Different line types correspond to the different redshifts shown in the caption of the Figure.}
\label{fig:corrfuncz}
\end{figure*}

The goal is to determine whether the preferential direction of infall found in the previous subsection at redshift zero is also valid at earlier redshifts, {\it i.e.}  if halos indeed form along a preferred axis. To this end, 1) particles that constitute the closely bound regions of the Virgo halos at redshift zero (i.e. particle within R$_{\rm 200}$) are selected, 2)  these particles are traced back to get their positions at an earlier redshift, 3) the Virgo halo progenitors at this same redshift are considered to obtain their center of mass and their R$_{\rm 200}$ radius. Then, the autocorrelation functions using the selected particles and their position at a given redshift and the center of mass and R$_{\rm 200}$ radius of the progenitors at the same redshift are derived as before.

Results are shown on Figure \ref{fig:corrfuncz}. As expected, particles closely bound at a given redshift (i.e. within R$_{\rm 200}$ of the halo progenitor) are homogeneously distributed  around the halo progenitors, regardless of the redshift considered. The trend is slightly different when considering all the selected particles: at earlier redshifts some of the particles, to be closely bound at redshift 0, are (still) infalling onto the halo progenitors. When considering only these soon to be closely bound particles, a clear trend of a preferential direction of infall onto the halo progenitors becomes visible. The non-uniformity of the distribution of infalling particles becomes clearer as the progenitors grow, i.e. as the redshift tends to zero. The direction of infall has appreciably but not drastically changed over time along with the growth of progenitors and the evolution of the Large Scale Structure (see Table \ref{Tbl:3}). 

\subsection{Merging history}

\begin{figure}
\hspace{-0.5cm}\includegraphics[width=0.52 \textwidth]{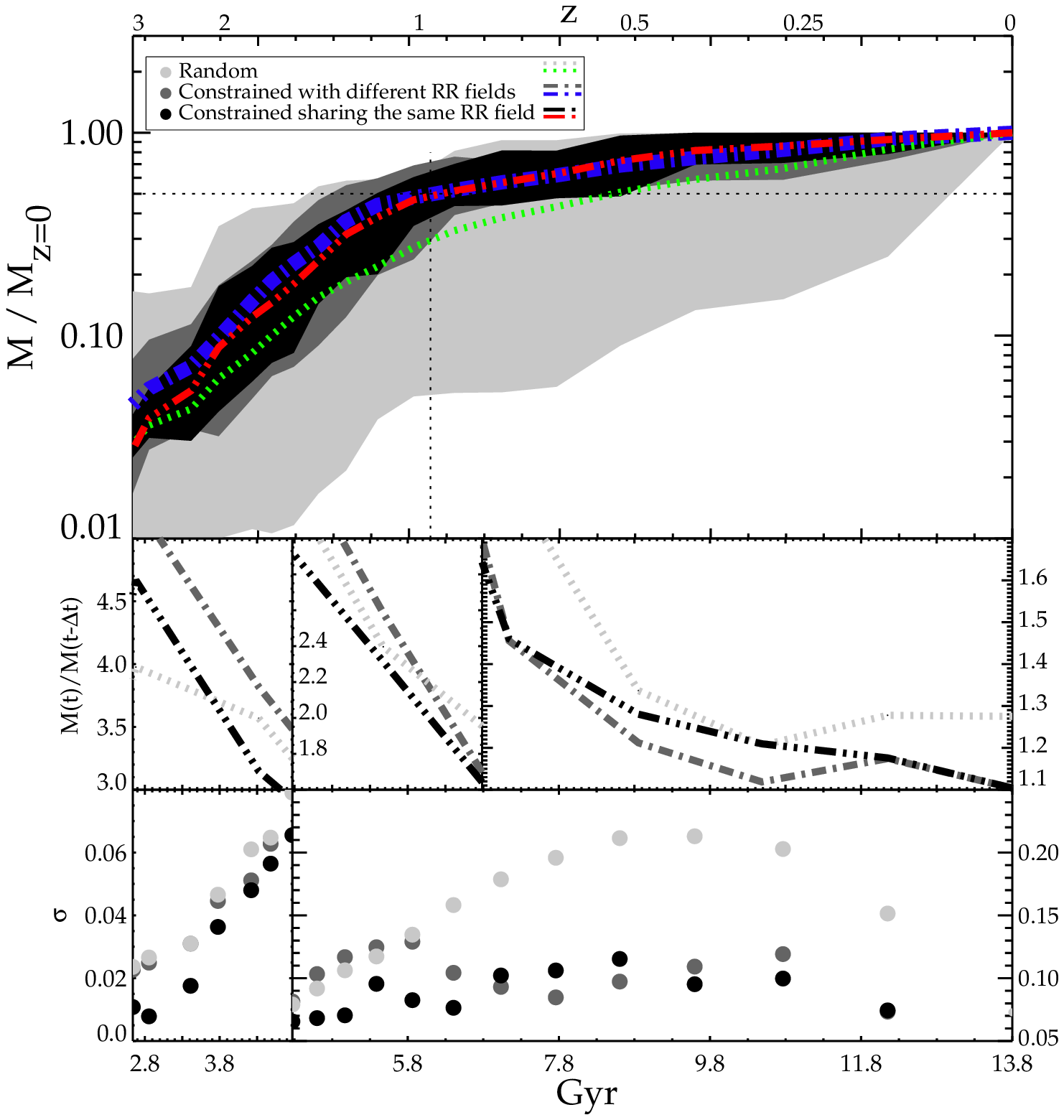} 
\caption{Top: Zones of possible merging histories obtained with 100 random halos sharing the same mass range as constrained halos (light grey), 15 Virgo halos from simulations built with different large scale random fields (dark grey) and 10 Virgo halos from simulations sharing the same large scale random field (black). The mean merging histories are plotted on top of the regions (green dotted line for random, blue dot-dashed line for constrained with different large scale random fields and red triple dot-dashed line for constrained sharing the same large scale random field). Masses at every redshift have been divided by the mass at redshift zero. Middle: ratios of masses at two different times such that two masses are separated by a constant time interval. The only difference between the three panels is the y-axis scale that varies to better show the trend with the redshift. Bottom: Standard deviation of the merging histories about their mean. Only the y-axis scale changes in the two panels. Middle and bottom panels share the same grey color code and the same linestyle as the top panel.}
\label{fig:merghist}
\end{figure}

Merging histories of the 25 Virgo halos are compared with those of 100 halos, selected randomly in the same mass range, at redshift zero, as the former.  At a given redshift, minimum and maximum mass of any progenitor are identified and the corresponding interval is plotted in the upper panel of Figure \ref{fig:merghist}. Three samples are considered: the 15 Virgo halos from the different-RR simulations (dark grey), the 10 Virgo halos from the same-RR simulations (black) and the 100 randomly selected halos (light grey). Therefore, these regions define the possible merging histories of the samples. All the masses are normalized to that at redshift zero. For each sample, the mean merging history is plotted on top of the corresponding area with blue dot-dashed, red triple dot-dashed and green dotted lines. The middle panel shows the accretion rate while scatters about the means are plotted in the bottom panel of Figure \ref{fig:merghist}. 

The 10 Virgo halos from the simulations sharing the same large scale random field in black present a narrower range of merging histories than the 15 Virgo halos of simulations built from different large scale random fields, in dark grey, as expected. The Virgo halos present a smaller scatter in their merging histories than the 100 unconstrained randomly selected halos spanning over the same mass range (light grey area in Figure \ref{fig:merghist}).  Clearly, at low redshifts, the variance of constrained halos' merging histories is decreased by a factor $\sim$ 2 when compared to that of the merging histories of random halos.  

Moreover, the mean merging histories differ for the random halos and for the Virgo halos: there is a break at redshift $\sim$ 1 in the mean merging history of Virgo halos, that is not visible in that of the random halos. The accretion of material onto halos becomes smoother with time for Virgo halos  than for an average random halo of the same mass. At approximately the same redshift, Virgo halos have acquired $\sim$50\% of their redshift zero masses while the average random halo has gathered only $\sim$ 30\% of its mass. Namely, the large scale environment of the Virgo cluster considerably constrains its possible evolution. \\

To evaluate the difference in accretion between the random halos and the Virgo halos, we define the following mass ratios
$M[t]/M[t-\Delta t]$ where $\Delta t$ is defined such that $t-\Delta t$ always represents the same amount of time (1.7 Gyrs). The middle panel of  Figure \ref{fig:merghist} shows the average ratios computed from the mean merging histories displayed in the top panel. Virgo halos tend to have higher accretion rates than random halos at high redshifts but as halos grow more massive, this  trend is reversed. Actually,  when the constrained Virgo halos have reached $\sim$ 50\% of their masses, an average random halo starts to accrete mass faster than Virgo halos and although the accretion rates keep decreasing for both Virgo halos and random halos, it still stays higher for an average random halo than for the Virgo halos. At redshift close to zero, a plateau is almost reached with rather small accretion rates  that are  nevertheless  larger for the average unconstrained halos  than for the constrained ones.


\section{Conclusion}

Using 25 constrained simulations of the evolution of the Local Universe, the formation history of the Virgo cluster is studied. 
More precisely, in each one of the 25 simulations that include 10 simulations built with the same large scale random field but different small scale features, a unique Virgo dark matter halo is found within less than 4 \hMpc\ from the observed Virgo cluster position. Parameters such as coordinates and distances, velocities and velocity dispersions are within 10-20\% of their observational counterparts and the different simulations give results in agreement at the same level. As expected the 10 simulations sharing the same large scale random field are among themselves in better agreement, the scatter reduces to  5-10\%. 

For a better comparison between observational and simulated masses, we rely on a method based on the zero-velocity radii (distance from the center of a system at which system-centric velocities of objects are zero). Often used by observers, it permits to derive mass estimates of clusters completely independent from the virial masses subject to large biases. In order to reproduce the observational method with the dark matter halos, in rough approximation dark matter particles are assumed to trace the galaxies and a Hubble flow is added to their radial peculiar velocities computed with respect to the observer in the center of the box. Then halo-centric velocities for the particles are derived and a spherical collapse model is fit in the resulting Hubble diagram. The intersection of the fit with the x-axis gives the zero-velocity radii and  masses are obtained by summing all particles up to this radius. For comparisons, the resulting masses of the Virgo halos are typically about 45\% (10\%) higher than M$_{\rm 200}$ (M$_{\rm vir}$) given by the Halo Finder and they are in excellent agreement (within one-sigma) with the most recent mass estimates obtained for the Virgo cluster in a similar manner \citep{2014ApJ...782....4K}.\\

Next a preferential direction of infall onto the Virgo halos is searched for. Particles infalling onto Virgo halos move along a preferred direction that is similar in all the simulations. Actually, this direction is along the filament in which the Virgo cluster resides, pointing towards coordinates similar to that of the Abell 1367 cluster as predicted by observations. It corresponds also to the direction of slowest collapse. This is in agreement with the Cosmic Web analyses according to which the clusters accrete matter along their host filament that appears to be aligned with the direction of slowest collapse.

Finally, the merging histories of Virgo halos have a cosmic variance reduced by a factor 2 at low redshifts with respect to those of random halos of the same mass: {\it i.e.}, merging histories of randomly chosen halos of the same masses as the Virgo halos span over twice a larger range of possible histories. Interestingly, at around redshift 1,  Virgo halos have already accreted 50\% of their mass while  an average random halo of the same mass has only accreted 30\% of its mass. This suggests a relatively quiet merging rate for the Virgo cluster during the last 7 Gyrs compared to a random cluster of the same mass. This knowledge may be of extreme importance for observational analyses.\\

In the 25 constrained simulations, 25 Virgo halos with properties consistent with those of the Virgo cluster are identified. The next step is to perform  a series of  zoomed simulations of the cosmic volume surrounding Virgo. Very high resolution Dark Matter only simulations combined with  semi-analytical models of galaxy formation and full hydrodynamical simulations will offer the possibility to compare the observed and simulated Virgo cluster in more details.

\section*{Acknowledgements}

The authors gratefully acknowledge the Gauss Centre for Supercomputing e.V. (www.gauss-centre.eu) for providing
computing time on the GCS Supercomputers SuperMUC at LRZ Munich and Jureca at JSC Juelich. JS acknowledges support from the Alexander von Humboldt Foundation. SG and YH acknowledge support from DFG under the grant GO563/21-1. GY acknowledges support from the Spanish MINECO under research grants AYA2012-31101 and AYA2015-63810.  YH has been partially supported by the Israel Science Foundation (1013/12). 

\section*{Appendix: Halo-centric velocities}

\begin{figure}
\centering
\includegraphics[scale=0.3,angle=-90]{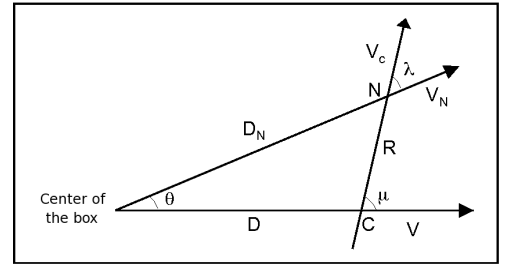}
\caption{Diagram representing the different parameters required to compute the distance and the velocity of a particle with respect to the center of a halo.}
\label{schema}
\end{figure}

To compute halo-centric velocity, the perspective of the outflow or perturbed Hubble Law is considered. With the help of Figure \ref{schema} and \citet{2006Ap.....49....3K}, let C be the center of a halo located at the distance D and receding with a velocity V from the center of the box where the observer is assumed to be. Consider a particle N at the distance D$_N$ and receding at V$_N$ from the center of the box. We call the angular separation, between the center of the system and the particle N, $\theta$.  Then, the distance between the particle N and the center of the system R$_c$ can be expressed as follows:
\begin{equation}
R_c=R=\sqrt{D^2 + D_N^2 - 2 D D_N cos \theta}
\end{equation}
The particle N is going away from C at the velocity:
\begin{equation}
V_c=V_N cos \lambda - V cos \mu
\end{equation} 
where $\mu = \lambda + \theta$ and tan $\lambda$ = $\frac{D sin \theta}{D_N - D cos \theta}$ assuming that random peculiar velocities are low compared to expansion velocities.

\clearpage

\bibliographystyle{mnras}

\bibliography{biblicomplete}

\end{document}